\documentclass[preprint,floatfix] {revtex4} 
\newcommand{\rvec}{\mathrm {\mathbf {r}}} 
\newcommand{\nvec}{\mathrm {\mathbf {n}}}

\newcommand{\Cvec}{\mathrm {\mathbf {C}}} 
\newcommand{\Svec}{\mathrm {\mathbf {S}}}

\newcommand{\Pvec}{\mathrm {\mathbf {P}}} 
\newcommand{\Uvec}{\mathrm {\mathbf {U}}}

\newcommand{\Hvec}{\mathrm {\mathbf {H}}} 
 
\newcommand{\Avec}{\mathrm {\mathbf {A}}}

\newcommand{\Evec}{\mathrm {\mathbf {E}}} 
\newcommand{\Qvec}{\mathrm {\mathbf {Q}}} 
\newcommand{\Dvec}{\mathrm {\mathbf {D}}} 
\newcommand{\vvec}{\mathrm {\mathbf {v}}}

\usepackage{color}
\usepackage{graphicx,booktabs}
\usepackage{graphics}
\usepackage{subfigure}
\usepackage{enumerate}
\usepackage{tikz}
\usepackage{amsmath}
\usepackage{amssymb}
\usepackage{bm}
\usepackage{tikz} 
\usepackage{xcolor}
\usepackage{enumitem}
\usepackage{epigraph}
\usepackage{filecontents}
\usepackage{rotating}
\usepackage{adjustbox}
\usepackage{mathptmx}

\begin{document}

\title{A real-time TDDFT scheme for strong-field interaction in Cartesian coordinate grid}

\author{Abhisek Ghosal\footnote{Present Address: Department of Chemical Sciences, Tata Institute of 
Fundamental Research (TIFR), Mumbai-400005, Maharastra, India}}
\author{Amlan K.~Roy}
\altaffiliation{Email: akroy@iiserkol.ac.in, akroy6k@gmail.com.}                                           
\affiliation{Department of Chemical Sciences\\
Indian Institute of Science Education and Research (IISER) Kolkata, \\  
Nadia, Mohanpur-741246, WB, India.}

\begin{abstract}
In this communication, we present a new approach towards RT-TDDFT through time-dependent KS equations 
based on an \emph{adiabatic eigenstate subspace} (AES) procedure. It introduces a second-order split 
operator technique in energy representation to implement the approximate TD propagator in AES. 
Most of the elements in TDKS matrix are directly computed in Cartesian coordinate grid (CCG). 
To demonstrate the internal consistency of our proposed scheme, we computed the TD dipole moment and 
high harmonic generation spectra using an adiabatic local density approximation. The comparison with available theoretical 
results ensures the feasibility of this proposed route. 


\vspace{5mm}

\end{abstract}
\maketitle

\section{Introduction}
Throughout the past several decades, time-dependent density functional theory (TDDFT) has been found to be the most popular 
method for computing linear response and excitation energies of many-electron systems, mostly on atoms, molecules, and 
molecular clusters, due to its reasonable trade-off between accuracy and efficiency \citep{casida12,marques12,adamo13}. It can 
be viewed as a reformulation of time-dependent (TD) quantum mechanics where the basic variable is TD single-particle 
electron density, $\rho(\rvec,t)$, instead of many-electron wave function. This is based on a one-to-one correspondence 
between TD external potential and $\rho(\rvec,t)$, for many-body systems evolving from a given initial state \citep{runge84}, 
what is called TDDFT. Now, the most widely recognized and implemented version of TDDFT is that of linear-response theory 
(LR-TDDFT) based on Casida's equation \citep{casida95}. It has been developed for computing excitation energies, mostly for 
discrete spectrum. In comparison to the conventional wave function-based formalisms, such as CI, MR-CI, complete active space 
self-consistent field, equation-of-motion coupled cluster etc., LR-TDDFT has found wide applicability in medium to large 
systems, for same obvious reason. This is evident from its applicability to medium to large systems \citep{casida12, marques12, 
adamo13}. Though it is found to be the most economically efficient method to compute low-lying excited states for finite 
systems, it is limited to weak-field processes (in which the density is perturbed slightly and remains close to ground state). 

\noindent
\par
As an alternative to LR-TDDFT in frequency domain, one can directly solve the time-dependent Kohn-Sham (TDKS) equation 
\citep{yabana96} (or 
its variant, the Liouville-von Neumann equation based on density matrix), the so-called \emph{real-time} TDDFT (RT-TDDFT). 
This approach is 
non-perturbative \emph{in principle}, and has the ability to capture all the non-linear effects in a photo-chemical and 
physical process. These include non-linear optical properties \citep{ding13}, photo-induced coupled electron-ion dynamics 
\citep{falke14} namely, photo-dissociation and photo-isomerisation, electronic quantum control, and strong-field 
processes such as above-threshold ionization \citep{schafer93}, high harmonic generation (HHG) \citep{chu01}, 
multi-photon ionization \citep{sakai03}, charge-resonance enhanced-ionization \citep{bocharova11}, etc. With the 
advancement in atto-second science and relevant spectroscopic techniques, one can now probe directly the electron dynamics 
with femto- or atto-second temporal resolution. That is why further development of theoretical as well as sophisticated 
computational methods are essential to complement these strong-field processes \citep{ullrich12, crawford14, nisoli17, 
kruchinin18}. Motivated by the rapid developments in ultra-fast experimental techniques, RT-TDDFT is drawing more and 
more attention as a reliable method of choice to simulate the strong-field phenomena. 

\noindent
\par
In order to solve the TDKS equations, one, \emph{in principle}, needs a full time-dependent propagator (TDP) which can be written as, 
\begin{equation}
\Uvec(t,0)=\mathcal{T} \mathrm{exp}\bigg\{-i\int_{0}^{t}dt^{\prime} H_{\mathrm{s}}(\rho,t^{\prime})\bigg\}, 
\end{equation} 
where $\mathcal{T}\mathrm{exp}$ is the time-order exponential containing TDKS Hamiltonian, 
$H_{\mathrm{s}}(\rho,t)$. But in practice, it has been used for a short time of interval, and then the 
full-TDP can be rewritten as,
\begin{eqnarray}
\Uvec(t_1,t_2)=\Uvec(t_1,t_3)\Uvec(t_3,t_2) \nonumber \\  
\Uvec(t,0)=\prod_{i=0}^{N-1}\Uvec(t_i+\Delta t_i,t_i).
\end{eqnarray}
In general, the TDP is unitary, $\Uvec(t+\Delta t,t)=\Uvec^{-1}(t+\Delta t,t)$, and has the time-reversal 
symmetry, $\Uvec(t+\Delta t,t)=\Uvec^{-1}(t,t+\Delta t)$, for a Hermitian Hamiltonian, which is obvious 
for $H_{\mathrm{s}}(\rho,t)$. As its exact form is non-trivial to implement as yet, it has to be 
approximated, and accordingly various \emph{approximate time-dependent propagators} (ATDP) have been introduced 
in the literature, such as exponential 
midpoint and modified midpoint algorithm, enforced time-reversal symmetry method, Magnus expansion, 
splitting techniques, etc \citep{castro04, li05, watanabe02}. Recently, the so called classical based ATDPs 
have been utilized in RT-TDDFT calculations which were largely ignored in the past \citep{gomez18}. This 
includes linear multi-step, Runge-Kutta, exponential Runge-Kutta and their different versions. The 
accuracy and efficiency of a RT-TDDFT calculation depends critically on the level of ATDP and spectral 
characteristic of TDKS Hamiltonian. These have been implemented successfully using plane waves 
\citep{schleife12}, numerical orbitals \citep{soler02, hekele21}, Gaussian-type orbitals \citep{li05, lopata11, 
nguyen15}, and on real-space grids \citep{andrade15} for a wide range of RT-TDDFT applications.
 
\noindent
\par
We, however, note that RT-TDDFT is not used widely as the method of choice in literature due to its huge 
computational burden requiring a smaller time step to propagate the density for a period of time (e.g., tens 
of femtoseconds at a minimum). On the other hand, there are no such alternatives which can accurately describe 
the non-linear electron dynamics of a large quantum mechanical system. Besides, high-lying excitations are 
attractive targets for RT-TDDFT which are very expensive for linear response algorithm due to the iterative 
nature of implementation. In recent years, most developments have been found to reduce the 
computational cost while maintaining the accuracy and numerical stability of ATDP. In this direction, we 
offer a novel propagation scheme based on \emph{adiabatic eigenstate subspace} (AES) technique that 
holds the promise of a larger time step. 
This advocates the use of a split-operator technique in energy representation
\citep{son09}, to perform the ATDP in AES. We have used a density predictor/corrector algorithm to 
maintain the self-consistency in $\rho(\rvec,t)$ at each time step. Most of the elements of TDKS matrix are 
directly computed in Cartesian coordinate grid (CCG). The relevant TD properties are computed through TD 
dipole-moment (TDDM). To demonstrate the internal consistency of our proposed scheme, we consider the interaction 
of a molecule with a high-intensity ultra-short strong laser field, taking H$_2$ and N$_2$ as representative.  
The TDDM and HHG spectrum are calculated using \emph{adiabatic local density approximation} (ALDA), along with a comparison 
of available theoretical references 
\citep{nwchem20}. This would establish the main objective of present communication to offer a general framework of 
RT-TDDFT of many-electron systems for strong-field processes.  
Towards this goal, we present a RT-TDDFT implementation in our in-house code InDFT \citep{indft19}.  
The plan of this article is as follows. In next section, we discuss the structure of TDKS 
equation and its representation in AES, along with the propagation scheme. Then, we highlight some of the TD quantities 
in CCG. In Sec.~III, the necessary computational and technical details are offered. In Sec.~IV, we demonstrate 
the internal consistency of this approach in strong-field 
regimes considering TDDM and HHG spectrum. For H$_2$, an \emph{all-electron} calculation is performed, whereas a 
\emph{pseudopotential} approximation is invoked in case of N$_2$. Finally, a few conclusions as well as future and outlook 
are summarized in last section.

\section{Methodology}
\subsection{Time-dependent KS equation}
Similar to ground state KS-DFT \citep{kohn65}, the TD non-interacting system is described by a single Slater determinant 
of fictitious orbitals ($\bm{\varphi}(t)=|\varphi_1(\rvec,t),\varphi_2(\rvec,t),\dots,\varphi_j(\rvec,t)|$) and the 
interacting TD electron density $\rho(\rvec,t)$ is represented as, 
\begin{equation}
\rho(\rvec,t) = \sum_{j=1}^{N} |\varphi_{j}(\rvec,t)|^{2},
\end{equation} 
where $N$ is the total number of occupied orbitals. Now, these 
orbitals are obtained by solving the TDKS equation as below (henceforth atomic units employed 
unless otherwise stated), 
\begin{eqnarray}
	i\frac{\partial}{\partial t}\varphi_{j}(\rvec,t) = \text{H}_{\mathrm{s}}(\rvec,t) \varphi_j(\rvec,t), \\
	\text{H}_{\mathrm{s}}(\rvec,t) = \big[-\frac{1}{2}\nabla^2 +v_{\mathrm{s}}(\rvec,t)\big], \nonumber \\
v_{\mathrm{s}}(\rvec,t)=v_{\mathrm{ext}}(\rvec,t)+v_{\mathrm{h}}(\rvec,t)+v_{\mathrm{xc}}(\rvec,t).  \nonumber
\end{eqnarray} 
Here $v_{\mathrm{s}}(\rvec,t)$ signifies the one-body TDKS potential. It includes the external potential, 
$v_{\mathrm{ext}}(\rvec,t)$, the classical Coulomb or Hartree potential, $v_{\mathrm{h}}(\rvec,t)$ and the 
TD exchange-correlation (XC) potential, $v_{\mathrm{xc}}(\rvec,t)$. We restrict ourselves 
to spin-unpolarized systems for simplicity.

\subsection{Adiabatic eigenstate subspace}
The TDKS Hamiltonian, $\text{H}_{\mathrm{KS}}$ consists of both linear and non-linear terms. The former including only 
the kinetic term, is considered for possible \emph{stiffness} of TDKS equation. Once the system is excited from
ground state, the TD density will oscillate with a wide range of frequencies. The Fourier components of those 
oscillations correspond to the Bohr frequencies, including both valence-to-valence and core-to-valence excitations. To 
maintain the numerical stability due to presence of such highly oscillating Fourier components, Eq.~$(4)$ has to be 
solved using a sufficiently small time step, which is the main integrated part of ATDP. Recently, AES approaches with
varying flavor have been introduced in the literature \citep{chu01, son09, russakoff16} to increase the time steps and 
hence reducing the computational cost. Accordingly, the TDKS orbitals can be expanded as,
\begin{eqnarray}
\varphi_{j}(\rvec,t) = \sum_{l} a_{jl}(t) \phi_l(\rvec), \\
a_{jl}(t) = \delta_{jl} \quad \mathrm{at} \quad t=0. \nonumber
\end{eqnarray}
where $\phi_l (\rvec)$ denotes the solution of time-independent KS (TIKS) equations at its ground state. This forms the basis of AES 
and the expansion coefficients, $a_{jl}$(t) projects the TDKS orbitals into this subspace.

\par
In order to utilize the advantage of AES, we recast the Hamiltonian in following way \citep{son09}, 
\begin{equation}
	\text{H}_{\mathrm{s}}(\rvec,t)=\text{H}_{\mathrm{s}}^{\mathrm{st}}+\text{H}_{\mathrm{s}}^{\mathrm{dy}}.
\end{equation}
where $\text{H}_{\mathrm{s}}^{\mathrm{st}}(\rvec)=-\frac{1}{2}\nabla^2 +v_{\mathrm{s}}(\rvec,0)$ and 
$\text{H}_{\mathrm{s}}^{\mathrm{dy}}(\rvec,t)=v_{\mathrm{ks}}(\rvec,t)-v_{\mathrm{ks}}(\rvec,0)$. The first 
one is called stationary KS Hamiltonian corresponding to ground state. This would remain 
constant throughout the propagation. The second one stands for dynamic KS Hamiltonian which effectively 
contains the fluctuation embedded in TD potential. It should be updated at each instant of time during 
propagation. This kind of splitting may help us to utilize the outcome of AES approach fully, and 
consequently may improve the computational efficiency.

\noindent
\par 
We are now able to explore the effect of AES on TDKS equations in a well-defined matrix form by exploiting 
Eqs.~$(4)$ to $(6)$. After some mathematical manipulation, we can recast Eq.~$(4)$ in terms of TD expansion 
coefficients in following fashion,
\begin{equation}
i \frac{\partial a_{jk}}{\partial t}  =  a_{jk}(t) \epsilon_k+ \sum_l a_{jl}(t) 
	\langle \phi_k|\text{H}_{\mathrm{s}}^{\mathrm{dy}}|\phi_l\rangle, 
\end{equation}
where $a_{jk}$ is the $k$th coefficient of $j$th TDKS orbital and $\epsilon_k$ denotes the eigenvalue of $k$th TIKS 
orbital from which the AES is formed. Further, we can also set up all the equivalent equations of the 
remaining coefficients for $j$th TDKS orbitals in a well-defined matrix form
\begin{equation}
i \frac{\partial \Avec_{j} }{\partial t} = \big(\bm{\epsilon} + \Hvec_{\mathrm{s}}^{\mathrm{dy}}\big) \Avec_{j}.
\end{equation}
Here, $\Avec_{j}$ stands for a column matrix for the $j$th TDKS orbital and $\bm{\epsilon}$ is a diagonal matrix 
containing the eigenvalues corresponding to TIKS orbitals. In general, $\Hvec_{\mathrm{s}}^{\mathrm{dy}}$ is a 
complex Hermitian matrix and, therefore, it must be square symmetric \citep{lopata11, zhu18}. Again, it can be easily 
generalized to the spin-polarized case, as given below, 
\begin{equation}
i \frac{\partial \Avec_{j,\sigma} }{\partial t} = \big(\bm{\epsilon}_{\sigma} + \Hvec_{\mathrm{s},\sigma}^{\mathrm{dy}}\big) 
\Avec_{j,\sigma}, \quad \quad \sigma = \alpha ~ \mathrm{or} ~ \beta ~ \mathrm{spin}.
\end{equation}
All the quantities are defined exactly in the same way as Eq.~$(8)$, but contain the appropriate spin part. 
Similarly, the total TD electron density would have contributions from both $\alpha$ and $\beta$ densities i.e., $\rho(\rvec,t) = 
\rho_{\alpha}(\rvec,t) + \rho_{\beta}(\rvec,t)$ and $\rho_{\sigma}(\rvec,t)=\sum_{j} |\varphi_{j,\sigma}(\rvec,t)|^{2}$. Now, 
Eq.~$(9)$ is the real workhorse in our present spin-polarized RT-TDDFT calculations.

\subsection{Time-evolution through split-operator}
In order to solve Eq.~$(9)$, we use a second-order split-operator technique in energy representation as an ATDP \citep{chu01, son09}. 
Accordingly, it can be expressed as,  
\begin{equation}
\Uvec_{\sigma}(t+\Delta t) = \bigg[\mathrm{exp}(-i\frac{\bm{\epsilon}_{\sigma}}{2}\Delta t)
\mathrm{exp}(-i\Hvec_{\mathrm{s},\sigma}^{\mathrm{dy}}\Delta t) \mathrm{exp}(-i\frac{\bm{\epsilon}_{\sigma}}{2}\Delta t)\bigg] 
+O(\Delta t^3),
\end{equation}
where the dynamic KS potential term appears sandwiched between two TIKS orbital energies. The reverse of this approach 
(KS orbital energy term sandwiched between two dynamical KS potential terms) is also equally legitimate, but it requires two times 
more computational cost than the former. It is to be noted that this form of ATDP would be unitary and unconditionally stable 
provided all the exponential terms are computed exactly. It provides a very reliable source of second-order method. Now, the 
solution of Eq.~$(9)$ can be expressed as, 
\begin{eqnarray}
\Avec_{j,\sigma}(t+\Delta t) = \Uvec_{\sigma}(t+\Delta t)\Avec_{j,\sigma}(t), \quad \nonumber \\
\Avec_{j,\sigma}(t+\Delta t) = \Evec_{\sigma} \Svec_{\sigma} \Evec_{\sigma} \Avec_{j,\sigma}(t), \quad \\
	\Evec_{\sigma} = \mathrm{exp}(-i\frac{\bm{\epsilon}_{\sigma}}{2}\Delta t) \quad \mathrm{and} \quad \Svec_{\sigma} = 
\mathrm{exp}(-i\Hvec_{\mathrm{s},\sigma}^{\mathrm{dy}} \Delta t). \nonumber
\end{eqnarray}
At first glance, Eq.~$(11)$ seems to be a chain of simple matrix-vector multiplication, but it involves exponential of 
$\bm{\epsilon}_{\sigma}$ and $\Hvec_{\mathrm{s},\sigma}^{\mathrm{dy}}$ in $\Evec_{\sigma}$ and $\Svec_{\sigma}$ respectively. The 
exponential of $\bm{\epsilon}_{\sigma}$ is trivially obtained by exponentiating every entry on the main diagonal. It becomes 
time-independent, if uniform temporal spacing is used throughout the propagation. On the other hand, it is not so trivial to 
obtain $\Svec_{\sigma}$, the exponential of $\Hvec_{\mathrm{s},\sigma}^{\mathrm{dy}}$. This can be performed using a broad range 
of techniques such as diagonalization, power series method, Lanczos algorithm, etc. Here, we have employed a spectral 
decomposition of $\Hvec_{\mathrm{s},\sigma}^{\mathrm{dy}}$ in AES (i.e., 
$\Hvec_{\mathrm{s},\sigma}^{\mathrm{dy}} = \Cvec_{\sigma} \Qvec_{\sigma} \Cvec^{\dagger}_{\sigma}$) and accordingly, it is found 
that,
\begin{equation}
\Svec_{\sigma}=\mathrm{exp}(-i \Hvec_{\mathrm{s},\sigma}^{\mathrm{dy}} \Delta t) = \Cvec_{\sigma} \mathrm{exp}(-i \Qvec_{\sigma} 
\Delta t) \Cvec^{\dagger}_{\sigma}.
\end{equation}
The diagonalization of $\Hvec_{\mathrm{s},\sigma}^{\mathrm{dy}}$ is straightforward to perform in AES through numerical routine, 
and that makes this split-operator approach unitary and stable. We, however, note that it is quite different from traditional 
split-operator techniques where $\Svec$ contains kinetic energy operator, usually solved in momentum space. 

\subsection{Time-dependent quantities} 
Now we discretize various quantities like TD electron density, TIKS and TDKS orbitals as well as dynamics KS Hamiltonian on a 3D 
simulation box having $x,y,z$ axes, as below,  
\begin{eqnarray}
r_{i}=r_{0}+(i-1)h_{r}, \quad i=1,2,3,....,N_{r}~, \quad r_{0}=-\frac{N_{r}h_{r}}{2}, \quad  r \in \{ x,y,z \},
\end{eqnarray}
where $h_{r}, N_r$ denote grid spacing and total number of points along each direction. The TD electron density 
$\rho(\rvec,t)$ in this grid may be simply written as (``g" symbolizes discretized grid),
\begin{equation}
\rho(\rvec_g,t) = \sum_{j=1}^{N} f_j |\varphi_{j}(\rvec_g,t)|^{2}, \quad \varphi_{j}(\rvec_g,t)= \sum_{l} a_{jl}(t) 
\phi_l(\rvec_g) \nonumber. 
\end{equation}

At this stage, the contributions of dynamic KS matrix are directly computed in CCG, 
\begin{eqnarray}
	\Hvec_{\mathrm{s},ij}^{\mathrm{dy}}=\langle \phi_i(\rvec)|\text{H}_{\mathrm{s}}^{\mathrm{dy}}(\rvec,t)|\phi_j(\rvec)\rangle \nonumber \\
	\Hvec_{\mathrm{s},ij}^{\mathrm{dy}}=h_x h_y h_z \sum_{g} \phi_i(\rvec_g)\text{H}_{\mathrm{s}}^{\mathrm{dy}}(\rvec_g,t)\phi_j(\rvec_g),
\end{eqnarray}
and subsequently total TD norm, $N(t)$ is also computed in CCG as, 
\begin{equation}
N(t) = h_x h_y h_z \sum_{g} \rho(\rvec_g,t).
\end{equation}
The construction of various TD potentials in $\text{H}_{\mathrm{s}}^{\mathrm{dy}}(\rvec_g,t)$ within the adiabatic approximation 
at a given time can be calculated analogous to their time-independent counterpart, and the details have been well documented in 
our earlier works 
\citep{roy08, roy08a, roy11, ghosal16, ghosal18, mandal19, ghosal19, roy21,roy21a}; hence not repeated here. Here we need to consider 
only the TD potential arising from an electric field. 

\begin{figure}[t]             
\centering
\includegraphics[scale=0.6]{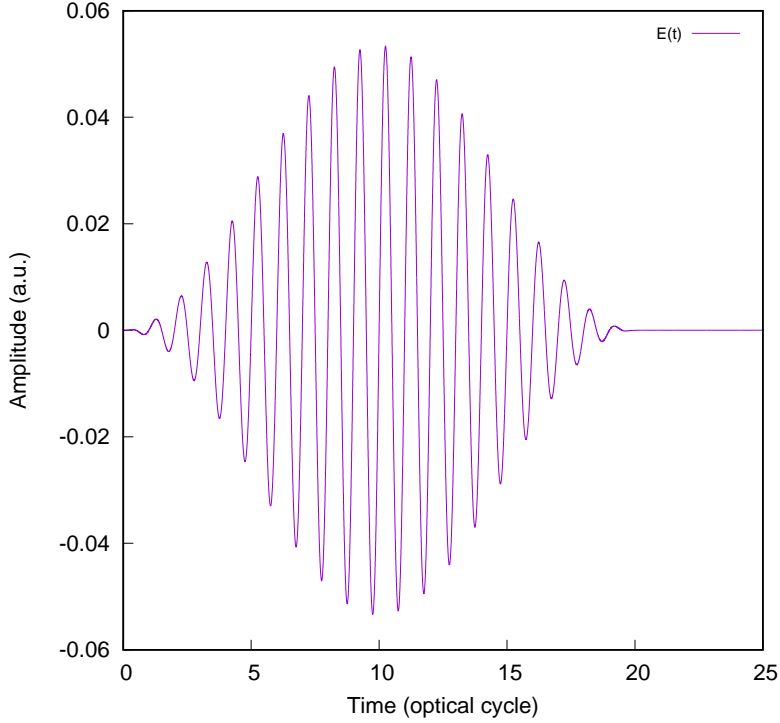}
\hspace{0.001in}
\caption{Electric field of the laser pulse with carrier frequency, $\omega_0=1.55$ eV and intensity, $I=10^{14}$ W cm$^{-2}$.}
\end{figure}

\noindent
\par
The TD electric field potential, $v_{\mathrm{f}}(\rvec,t)$ is introduced to represent the laser-matter interaction; 
in particular we are interested in laser-molecule interaction. It is considered to be a TD perturbation in 
$v_{\mathrm{ext}}(t)$ besides Coulombic attractive potential. During such interaction, the system is 
excited by applied electric field, $\Evec(t)$ through transition dipole moment integral. Within the ``length gauge" 
(or dipole) approximation, the component of $\Hvec_{\mathrm{s}}^{\mathrm{dy}}$ due to $v_{\mathrm{f}}(\rvec,t)$ can 
be expressed as,
\begin{eqnarray}
v_{\mathrm{f}}(\rvec,t)=-\Evec(t) \boldsymbol{\cdot} \rvec, \nonumber \\
\vvec_{\mathrm{f},ij}=-\bigg[\Dvec_{ij} \boldsymbol{\cdot} \Evec(t)\bigg],
\end{eqnarray}
where $\Dvec$ stands for transition dipole moment integral in AES. Here, we have mainly considered those properties 
which are derived from TDDM, $\mu(t)$. This can be computed directly from the transition dipole moment 
integral and TD density matrix, $\Pvec(t)$. Accordingly, the $z$-polarized component can be expressed as, 
\begin{equation}
\mu_z (t) = \mathrm{Tr}[\Dvec_{z}\Pvec(t)].
\end{equation}
The elements of $\Dvec_z$ in AES can be written as follows, 
\begin{eqnarray}
\Dvec_{z,ij} = \langle \phi_i(\rvec)|z|\phi_j(\rvec) \rangle, \nonumber \\
\Dvec_{z,ij} = \sum_{\nu \lambda} C_{\nu i} C_{\lambda j} 	
\langle \chi_{\nu}(\rvec)|z|\chi_{\lambda}(\rvec) \rangle,
\end{eqnarray}
where the first expression is further expanded in atomic orbital bases (or contracted bases) due to the fact that 
each element involved in Eq.~$(18)$ can be computed analytically using well-known Obara-Saika recursion relation 
\citep{obara86}. Moreover, the element of $\Pvec(t)$ in AES can be expressed in the following way,
\begin{equation}
\Pvec_{ij}(t) = \sum_{k} f_k a_{ik}(t) a_{jk}(t),
\end{equation}
where $a_{ik}$ is the $i$th coefficient of $k$th TDKS orbital.  

\begin{figure}[hbt!]             
\centering
\includegraphics[scale=0.6]{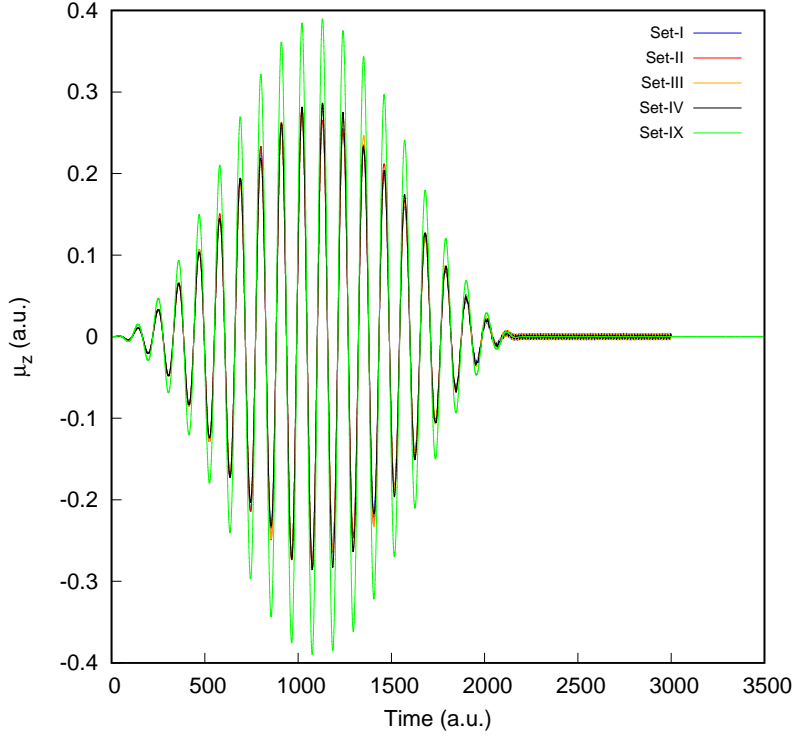}
\hspace{0.01in}
\caption{The temporal dependence of TDDM along z-direction of H$_2$ molecule over the 
	full duration of laser pulse, along with reference result (Set~IX).}
\end{figure}

\color{red}Before passing, it may be worth noting that the propagation scheme is not restricted to TIKS orbitals. 
One can use different subspaces consisting of atomic orbitals, numerical orbitals etc., but that may require additional cost 
during diagonalization of $\text{H}_{\mathrm{s}}^{\mathrm{st}}$ on that subspace. Here, our approach facilitates the  
removal of high-frequency oscillations generated by static Hamiltonian through the split-operator technique in energy 
representation. In this direction, it is worthwhile to mention the work of \citep{jia18} which is based on 
a parallel transport gauge transformation of orbitals such that the transformed orbitals, more specifically the 
residual vectors, are expected to oscillate less frequently than the original ones. But they do need additional term in the 
Hamiltonian in order to satisfy the desired transformation. On the other hand, the present approach is rather simple and it 
works because the spurious high frequencies of oscillation are damped by the $\mathrm{exp}(-i\frac{\bm{\epsilon}_{\sigma}}{2}\Delta t)$
term. Therefore the electron dynamics always follows the slowest possible (physical) oscillation.

\color{black}
\noindent
\section{Computational details}
For laser-molecule interaction, the electric field, $\Evec(t)$ is conveniently written as,  
\begin{equation}
\Evec(t)=\nvec f(t) \ \mathrm{sin}(\omega_{0}t),
\end{equation}
where $\nvec$ is the unit vector in the direction of electric field polarization vector, $f(t)$ is the envelope function; 
$\omega_0$ is the carrier frequency. Here, we employ a $\mathrm{cos}^2$ shape laser field, as below,  
\[
    f(t)= 
\begin{cases}
    f_0 ~ \mathrm{cos}^2 \left(\frac{\pi}{2\kappa}(\kappa-t) \right) & \text{if } |t-\kappa|\leq \kappa,\\
    0              & \text{otherwise}.
\end{cases}
\]
where $\kappa$ signifies the width of envelope. We choose the particular laser field with carrier frequency, $\omega_0=1.55$ eV 
(corresponding to a Ti:sapphire laser), intensity, $I=10^{14}$ W cm$^{-2}$, and $\kappa = 10$ optical cycle (o.c.), where $1$ 
o.c. $=\frac{2\pi}{\omega_0}$. Now, this particular form of laser parameters are adopted from \citep{luppi12} 
because it permits us to neglect the ionization effects. Then, the propagation is continued up to a maximum duration of $40$ 
o.c., corresponding to t$_f=4412.80$ a.u. ($106.74$ fs), where pulse duration is $20$ o.c. (see Fig.~$1$). 

\noindent
\par
The spectrum of HHG is computed by taking the Fourier transform of the TDDM:
\begin{equation}
H_z(\omega)=\Bigg|\frac{1}{t_{f}-t_i}\int_{t_i}^{t_f}\mu_{z}(t)e^{-i\omega t} dt\Bigg|^2.
\end{equation}
We consider the applied TD field along the molecular axis ($z$-direction) and the corresponding induced TDDM 
is measured along the same direction. In order to perform discrete Fourier transform, standard FFTW3 package 
\citep{fftw05} was invoked. 

\noindent
\par
The dimension of $\Svec$ is $N_{b,\mathrm{AES}}$ where $N_{b,\mathrm{AES}}$ defines the total number of eigenstates in AES. 
The computational cost of evaluation of exponential via diagonalization scales as $N_{b,\mathrm{AES}}^{3}$. It incurs 
a negligible cost compared to $\Hvec_{\mathrm{s}}^{\mathrm{dy}}$ formation, scaling as $N_{b,\mathrm{AES}}^{2}N_g$, where 
$N_g$ defines the total number of grid points in CCG. Moreover, Eq.~$(11)$ involves a series of matrix-vector 
multiplication scaling as $N_{b,\mathrm{AES}}^{2}$. Hence, construction of dynamic KS matrix remains the most 
computationally expensive step than any other operations. Further, it also requires twice the memory footprint of
ground-state DFT calculation as it becomes complex. Therefore, the numerous build of dynamic KS matrix may enhance the 
computational cost. In order to overcome this issue, we have used density predictor/corrector algorithm additionally for 
improving the efficiency in terms of number of $\Hvec_{\mathrm{s}}^{\mathrm{dy}}$ builds per unit time of propagation.
This is performed self-consistently by employing a convergence criteria during time evolution i.e., 
$\mathrm{max}\{|\rho^{\mathrm{new}}_{\sigma}-\rho_{\sigma}|\} < \iota$, where $\iota$ is about $1 \times 10^{-7}$. The 
resulting generalized matrix-eigenvalue problem and matrix diagonalization is solved through standard LAPACK routine 
\citep{anderson99} accurately and efficiently. These are implemented in our in-house program InDFT \citep{indft19}; 
all the results presented in this communication are computed from there.

\begingroup                      
\squeezetable
\begin{table}      
	\caption{\label{tab:table2} Numerical details of our proposed RT-TDDFT scheme in CCG (all in a.u.). See text 
for details.}
\centering
\begin{ruledtabular}
\begin{tabular} {lccccc}
Set & $h_r$ & $\Delta t$  & Basis set & XC functional & $\{ N_x, N_y, N_z \}$(optimal)  \\ 
\cline{1-6}
I & 0.2 & 0.2 & 6-311G & L(S)DA & $\{80, 80, 80\}$  \\
II & ,, & 0.1 & ,, & ,, & ,, \\
III & ,, & 0.05 & ,, & ,, & ,,  \\
IV & ,, & 0.025 & ,, & ,, & ,,  \\
V & 0.3 & 0.1 & SBKJC & ,, & $\{52, 52, 60\}$  \\
VI & ,, & 0.08 & ,, & ,, & ,,  \\
VII & ,, & 0.04 & ,, & ,, & ,,  \\
VIII & ,, & 0.01 & ,, & ,, & ,,  \\
\end{tabular}
\end{ruledtabular}
\end{table}
\endgroup

At first, we solve the TIKS equations, for a given basis set and XC functional. The corresponding TIKS orbitals form the AES 
which lies at the heart of our propagation scheme. A systematic grid optimization technique from \citep{ghosal16, ghosal18, 
mandal19, ghosal19, roy21, roy21a}, provides an optimal grid which would be the starting point of our RT-TDDFT calculations. It is 
well justified in terms of current propagation scheme. We have performed the calculations at fixed experimental 
geometry, taken from NIST database \citep{johnson16}, employing the following finite atomic orbital basis sets: 
\emph{all-electron} basis $6$-$311$G (includes only $s$-type of functions) for H$_2$ and ECP basis SBKJC \citep{stevens84} for 
N$_2$ respectively. For \emph{all-electron} calculations, we chose the particular one to avid any node 
near the nucleus that would facilitate the successful implementation of KS-DFT in CCG. These are adopted from EMSL Basis Set 
Library \citep{feller96}. For TDXC functional, we employ spin-polarized LDA in its ground state form within adiabatic 
approximation including Slater spin exchange and Vosko-Wilk-Nusair (VWN)-V parametrized model for correlation within a homogeneous 
electron gas approximation \citep{vosko80}. 

\begin{figure}[hbt!]             
\centering
\includegraphics[scale=0.60]{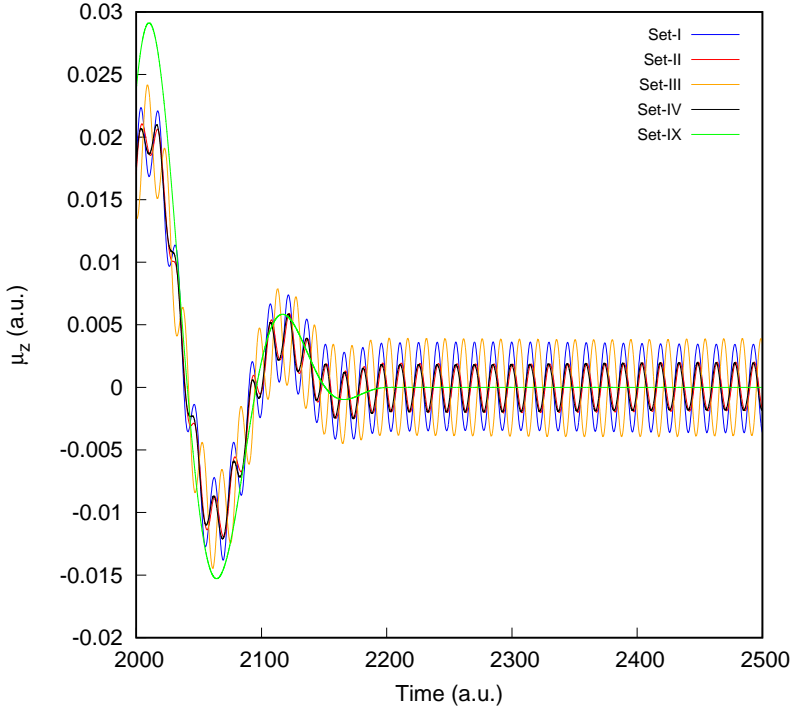}
\hspace{0.001in}
\caption{The temporal dependence of TDDM along z-direction of H$_2$ molecule toward the peak duration of pulse, 
	along with the reference result (Set~IX). See text for details.}
\end{figure}

\section{Results and Discussion}
In the present context, the primary objective is to introduce a simple CCG-based RT-TDDFT scheme in dealing with 
laser-molecule interaction, where perturbative models are not feasible. To the end, we would also discuss 
the effectiveness of pseudopotential RT-TDDFT calculation in simulating strong-field processes through a specimen 
case of N$_2$. Let us begin with the case of H$_2$ molecule. At first, the computed TDDM along $z$ direction is displayed 
in Figs.~2 and 3, for different temporal resolution. The first one gives it for full duration of the simulation, while 
the second one provides it towards the end of the pulse respectively. We project out the effect of temporal 
resolution on the accuracy by comparing them with reference results. The latter is identified as Set~IX, referring to 
6-311G basis set, spin-polarized LDA XC functional, $\Delta t=0.1$ and other default parameters in NWChem \citep{nwchem20}. 
The details regarding optimal grid, spatial resolution (or grid spacing), temporal resolution, $\Delta t$ are given in Table~I
for convenience. The polarization of the field was chosen to be the $z$-axis, along molecular axis. 

\begin{figure}[hbt!]             
\centering
\includegraphics[scale=0.60]{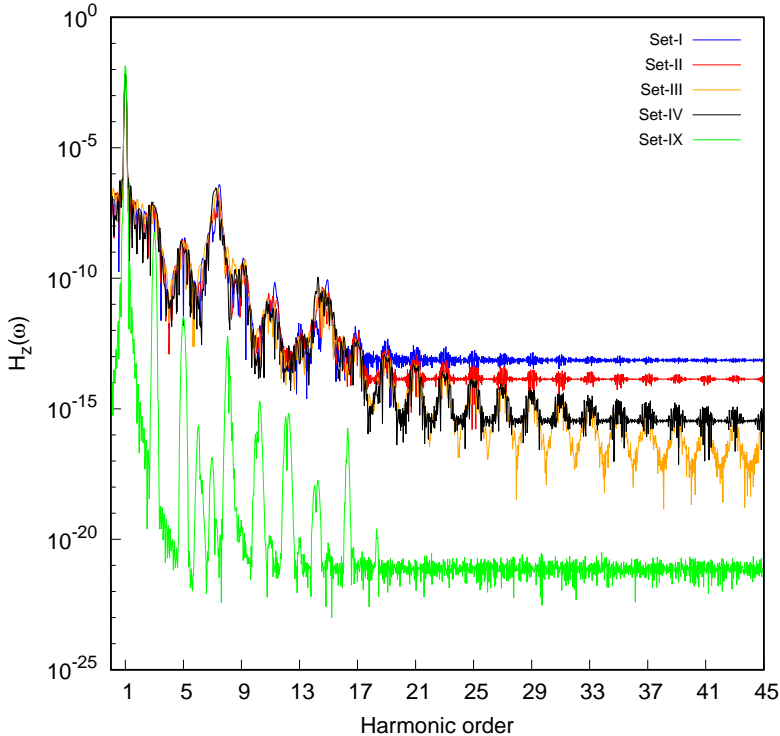}
\hspace{0.001in}
\caption{The temporal dependence of HHG spectra along z-direction of H$_2$ molecule compared with reference 
	result (Set~IX). See text for details.}
\end{figure}
 
\noindent
\par
We begin our computation with a temporal resolution of $0.2$ a.u., in Set~I which is twenty times larger than the
one used in \citep{luppi12}. This is then halved successively from Set~II to Set~IV. From Fig.~$2$, it 
is well manifested that TDDM coherently follows the $20$ o.c. of laser pulse 
irrespective of the set of interest, which corroborates well with \citep{luppi12}. Further, they also correctly 
follow the behavioral pattern of TDDM from reference (Set~IX), but the amplitude of oscillation at different peak 
positions are somehow different. Figure~3 tells that, the oscillation amplitude
at the end of $20$ o.c. does not return back to zero. It is quite possible due to non-adiabatic 
excitations during the propagation, as discussed in \citep{schlegel07}, but in our current scenario, we do not 
have enough flexibility in basis set to describe these excitations. We, however, believe that this small 
fluctuation in amplitude is due to numerical errors accumulated during the propagation. To analyze it 
further, we follow Fig.~$3$ very closely and it is observed that there is a sharp decrease in amplitude of 
oscillations than the frequency of laser field, and remains constant after Set~II. Accordingly, we set 
the time step corresponding to Set~III as an optimal choice for comparing the results with reference (Set~IX).

\begin{figure}[hbt!]             
\centering
\includegraphics[scale=0.60]{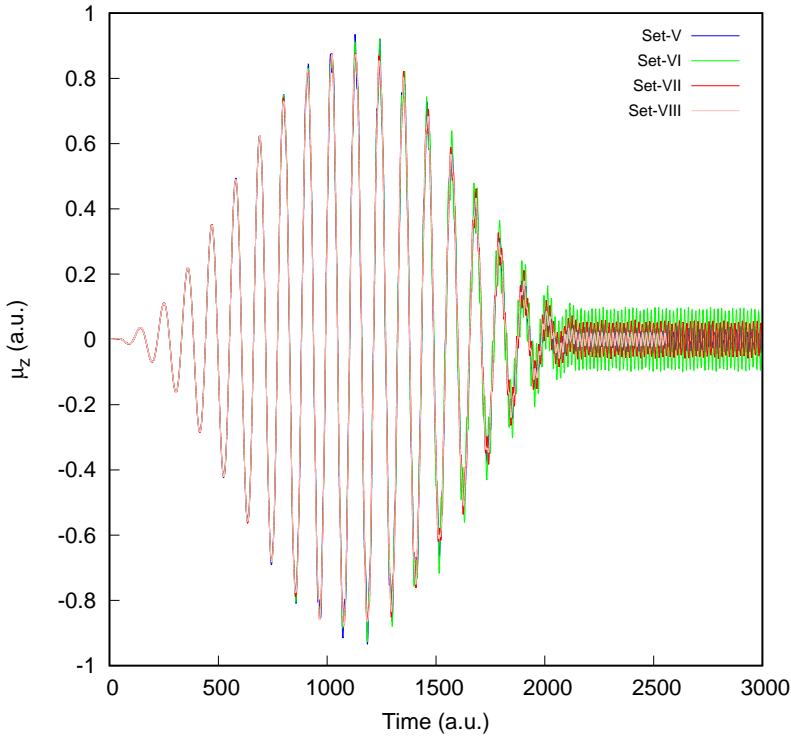}
\hspace{0.001in}
\caption{The temporal dependence of TD dipole moment along z-direction of N$_2$ molecule over the 
	full duration of the laser pulse.}
\end{figure}

\noindent
\par

Next, we present the HHG spectra from TDDM, in Fig.~$4$. Clearly, the distinctive 
shape of HHG spectra is obtained, which consists of a characteristic shape: first a rapid decrease for the 
low-order harmonics, followed by an extended plateau region, and finally a cut-off region. As expected, only odd 
harmonics are observed, because a homo-nuclear diatomic like H$_2$ has a center of inversion. All the HHG spectra 
have roughly comparable 
intensity profiles as expected from dipole. If we compare these spectra (Sets~I-IV), with reference (Set~IX), 
the intensity profiles are different, but all contain the same physics. The effects of smaller time step, as 
observed in dipole moment plot, are also visible in the characteristic of cut-off region. The results of Sets~I and II 
do not correctly reproduce the cut-off region of the spectra. To pursue further, we have calculated Keldysh parameter 
\citep{keldysh65} 
theoretically for the given basis set and XC functional, which is $0.93$ and the corresponding maximum harmonic 
cut-off is $22$. At this point, it is important to note that the calculated value of Keldysh parameter and maximum HHG 
cut-off is based on the assumption of a single active electron in the three-step model. 
It is well manifested by Sets~I and II, having a close resemblance with reference result (Set~IX). 
But for Sets~III and IV, the cut-off position moves towards the higher harmonics, which is also found in recent 
real-time wave function based calculation of H$_2$ \citep{luppi12}. We believe that the 
performance of our current calculations using spin-polarized LDA XC functional along with small basis set such as 
$6$-$311$G in describing HHG spectra may further be improved.  

\begin{figure}[hbt!]             
\centering
\includegraphics[scale=0.60]{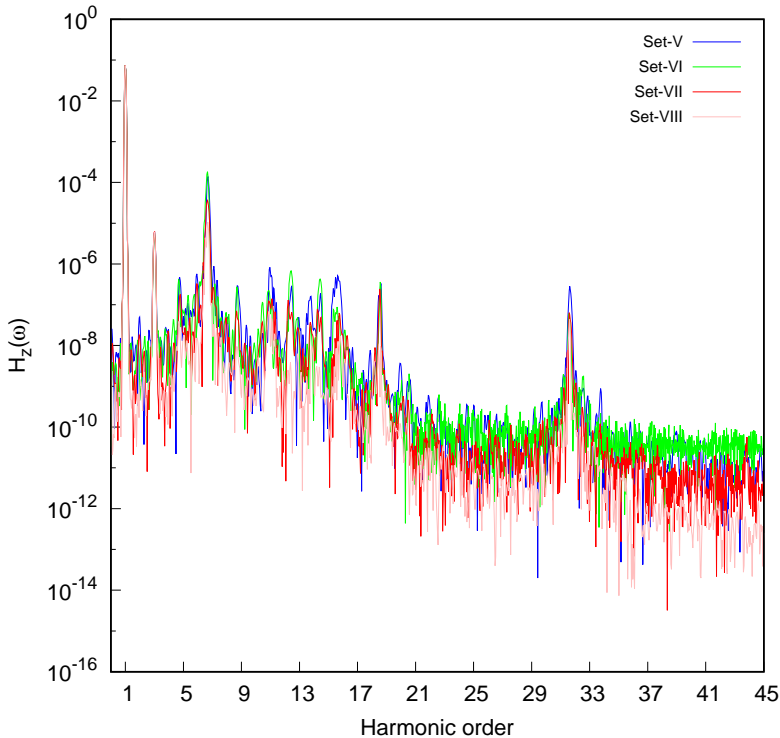}
\hspace{0.001in}
	\caption{The temporal dependence of HHG spectra along z-direction of N$_2$ molecule.} 
\end{figure}

\noindent
\par
In order to analyze the stability of propagator and its dependence on time step, we note that the norm 
of wave function was conserved in all Sets (I-IV) during the propagation. Further, the exponential density 
predictor/corrector algorithm took mostly two steps (sometimes three) to facilitate a larger time, 
maintaining the accuracy and numerical stability. Moreover, it may be possible 
to resolve the numerical artefacts by controlling spatial resolution to a certain extent, but it has an adverse effect 
on computational cost, due to an increase in number of grid points. 

\noindent
\par
In a broader context, the focal point of this work is to deal with larger systems and this is facilitated by 
the use of ECP in the realm of gaussian basis set. The pseudopotential approximation would be a useful 
option for successful implementation of RT-TDDFT in CCG. In such calculations \citep{foglia17}, this may
offer substantial advantages: (1) it reduces the dimension of TDDM, which may significantly alleviate the cost of 
propagation scheme (2) it also increases the value of required time-step, by eliminating the need to integrate the 
motion of inner electrons, associated with high-energy transition and fast oscillating frequencies. Thus, the 
relevance of pseudopotential approximation may be significant when dealing with heavy atoms, not only because of a reduction 
of dimension of TDDM, but also through a considerable decrease in propagation time. 

Towards this direction, we choose N$_2$ molecule. At first, computation is performed with a temporal resolution of $0.1$ a.u., 
with Set~V which is ten times larger than the one used in \citep{luppi12}, and then it is gradually decreased from Set~VI to 
Set~VIII. From Fig.~$5$, it is clear that TDDM coherently follows the $20$ o.c. of laser pulse 
for all the sets of interest. Also Fig.~$5$ shows the same trend at the end of $20$ o.c., as found in case of 
H$_2$. Again, here also, the conclusion remains same and it is possibly due to some numerical 
artefacts accumulated during propagation. Furthermore, there is a sharp decrease in amplitude of 
oscillations than the frequency of laser field. We have calculated the HHG spectra from TDDM, 
which is reported in Fig.~$6$. The distinctive shape of HHG spectra in low-order harmonics and plateau region 
is well obtained, but the sharp cut-off region is missing. In a closer look, odd harmonics are there in the lower 
and plateau region. All the HHG spectra have roughly comparable intensity as expected from TDDM, and 
the dependence of time step is not so prominent here. Moreover, the numerical noise present in spectra is very 
hard to remove after filtering through window function, and that causes considerable complexity in the analysis of spectra. 
At this stage, it is difficult to distinguish the structure and signal in the HHG spectra from the background. We have 
calculated Keldysh parameter \citep{keldysh65} theoretically for the given basis set and XC functional, which is 
$1.04$ (tunneling ionization) and the corresponding maximum harmonic cut-off is $21$. It is not sharply manifested 
in all the sets. In a recent article \citep{foglia17}, it has been shown that RT-TDDFT calculations for absorption spectra with 
pseudopotential approximation can be quite sensitive to the choice of basis function. As most of the basis sets 
designed for ECP are typically constructed to reproduce ground-state properties, this may not be as flexible for TDDFT 
calculations, which inherently requires both ground and excited states. Currently, we are looking into it. 

\color{red}
A few remarks may be made at this stage. The present work is a continuation of our ongoing work to establish a viable alternative 
for DFT calculations, for small to medium systems, through a CCG based pseudopotential KS-DFT framework using LCAO-MO ansatz as 
implemented in InDFT \citep{indft19}. Here, we have made an attempt to establish a general framework of RT-TDDFT for atoms/molecules 
allowing reasonable trade-off between accuracy and efficiency. Therefore, it is our intention to offer the performance of our  
scheme in CCG. It is to be noted that the present propagation scheme is not restricted to CCG. One can
use atom-centered grid such as Lebedev grid \citep{lebedev75, lebedev76} with Becke’s partitioning \citep{becke88} to construct the 
TD Hamiltonian in accordance with the modern computational code. However, as evident from the previous discussion, the applicability of 
the present approach is crucially dependent on various factors, especially basis set; however the present propagation scheme remains 
quite reliable and accurate. Thus it is our firm belief that the current propagation scheme is much more general than its current mode 
of implementation, and has considerable promise for electron dynamics of strong-field processes for atoms/molecules.
\color{black}
\section{Future and outlook}
We have demonstrated the viability and suitability of an AES based RT-TDDFT approach for electron dynamics in strong-field 
processes, through both all-electron and pseudopotential approximation. This was applied for H$_2$ and N$_2$ as prototypical cases. 
The properties derived from TDDM were offered. Besides these, their performances on different time steps were also presented. 

\noindent
\par
The most important conclusion is that in finite systems, it shows considerable promise for 
electron dynamics of strong-field processes for larger systems. The success of this approach relies on accurate 
implementation of our numerical propagator in CCG as well as estimation of all dynamical KS matrix elements. We primarily 
focused on the dependence of time step on TDDM and its corresponding HHG spectra, for which very limited 
information is available in CCG, so far. 

\noindent
\par
The added flexibility of basis set is very important here. So far, we have taken in to account only a particular kind of basis 
set designed for ECP calculations. But, other basis sets derived from diffusive quantum Monte-Carlo, may also throw some light 
on the quality of these spectra. This issue leaves ample room for future code improvement. 

\section{Availability of data}
Data available on request from the authors.

\section{Acknowledgement}
AG is grateful to UGC for a senior research fellowship. AKR sincerely acknowledges funding from DST SERB, New Delhi, India 
(sanction order: CRG/2019/000293). Partial financial assistance from BRNS, Mumbai (sanction order: 58/14/03/2019-BRNS/10255)
is acknowledged.  A special thank goes to Dr. N. Govind for numerous valuable and constructive comments. AG thanks Prof. Vikas and 
Dr. E. Luppi for useful discussions.

\bibliography{dftbib1.bib}

\end{document}